\documentclass[
preprint,
nofootinbib,
amsmath,amssymb,
aps,
]{revtex4-2}
\usepackage[colorlinks]{hyperref}
\usepackage{graphicx}
\usepackage{dcolumn}
\usepackage{bm}
\usepackage{hyperref}
\begin{document}
	\title{Matter vs Vacuum Oscillations in Atmospheric Neutrinos}
	\author{Jaydeep Datta}
        \email{jaydeep.datta@gmail.com}
	\affiliation{Saha Institute of Nuclear Physics, Bidhannagar, Kolkata 700064, India}
	\affiliation{Homi Bhabha National Institute, Anushakti Nagar, Mumbai 400094, India}
	
	\author{Mohammad Nizam}%
	\email{mohammad.nizam@tifr.res.in}
	\affiliation{Tata Institute of Fundamental Research, Mumbai 400005, India}%
	\affiliation{Homi Bhabha National Institute, Anushakti Nagar, Mumbai 400094, India}
        
	\author{Ali Ajmi}
        \email{ali.ajmi.3c@kyoto-u.ac.jp}
	\affiliation{Department of Physics and Astronomy, Kyoto University, Kyoto 606-8502, Japan}%
	
	\author{S.Uma Sankar}
        \email{uma@phy.iitb.ac.in}
	\affiliation{Department of Physics, Indian Institute of Technology Bombay, Mumbai 400076, India}%

	\date{\today}
\begin{abstract}

At present, both vacuum oscillations as well as matter modified oscillations can explain the data of 
atmospheric neutrino experiments and the long baseline accelerator neutrino experiments equally well. Given the 
important role the matter effects play in the determination of CP violation in neutrino oscillations,
it is imperative to establish the signal for matter effects unambiguously. In this work, we study the ability of
ICAL at INO to make a distinction between vacuum oscillations and matter modified oscillations. We find that it is possible to 
make a $3~\sigma$ discrimination in ten years, whether the atmospheric mass-squared difference is positive or negative.
\end{abstract}

\maketitle

\section{Introduction}

The pioneering water Cerenkov detectors, IMB \cite{Casper:1990ac,BeckerSzendy:1992hq} and Kamiokande \cite{Hirata:1992ku,Fukuda:1994mc}, observed the up down asymmetry of atmospheric muon neutrinos and established atmospheric neutrino anomaly. The next generation experiment, Super-Kamiokande \cite{Fukuda:1998mi}, measured the zenith angle dependence of atmospheric muon and electron neutrinos and established atmospheric neutrino oscillations. The initial analysis of Super-Kamiokande data was done using the hypothesis of two flavour vacuum oscillations, which provided a good fit to the data. It also has determined the magnitude of the corresponding mass-squared difference $|\Delta_{\rm atm}| \simeq 3 \times 10^{-3}$ eV$^2$ (but not its sign) and the mixing angle $\sin^2 2 \theta_{\rm atm} \simeq 1$. This determination played an important role in the design of the long-baseline accelerator neutrino experiments T2K \cite{Itow:2001ee} and NO$\nu$A \cite{Ayres:2007tu}. 

Evidence for neutrino oscillations also came from solar neutrino experiments \cite{Cleveland:1998nv,Hampel:1998xg,Abdurashitov:2009tn,Altmann:2005ix,Abe:2016nxk,Aharmim:2011yq}. Analysis of the solar neutrino data yielded the result $\Delta_{\rm sol} \sim 10^{-4}$ eV$^2$ and $\sin^2 \theta_{\rm sol} \sim 0.3$. The three known light neutrino flavour states \cite{Steinberger:1990hr} can mix to form three mass eigenstates with masses $m_1$, $m_2$ and $m_3$.  From these, we can define two {\it independent} mass-square differences, $\Delta_{21} = m_2^2 - m_1^2$ and $\Delta_{31} = m_3^2 - m_1^2$. Without loss of generality, we can identify $\Delta_{\rm sol} = \Delta_{21}$ and $\Delta_{\rm atm} = \Delta_{31} \approx \Delta_{32}$. The unitary mixing matrix, connecting the flavour eigenstates to the mass eigenstates, is called the PMNS matrix \cite{Maki:1962mu,Bilenky:1978nj}. In a manner similar to the quark mixing matrix \cite{Kobayashi:1973fv,Kuo:1986sk}, it is parameterized in terms of three mixing angles, $\theta_{12}$, $\theta_{13}$ and $\theta_{23}$, and a CP violating phase $\delta_{\rm CP}$. The data from the CHOOZ reactor neutrino experiment leads to the strong constraint $\theta_{13} \leq 10^\circ$ \cite{Apollonio:1997xe,Narayan:1997mk}. The smallness of this mixing angle implies that the solar and the atmospheric neutrino anomalies can be analyzed as independent problems within three flavour oscillation framework \cite{Narayan:1997mk}. It also leads to the identification $\theta_{\rm sol} \simeq \theta_{12}$ and $\theta_{\rm atm} \simeq \theta_{23}$.   

The solar neutrinos, produced at the core of the sun, undergo forward elastic scattering as they travel through the solar matter. This scattering leads to matter effect \cite{Wolfenstein:1977ue,Wolfenstein:1979ni,Mikheev:1986wj}, which modifies the solar electron neutrino survival probability ($P_{\rm ee}$). Super-Kamiokande \cite{,Abe:2016nxk} and SNO \cite{Aharmim:2011yq} have measured $P_{\rm ee}$ as a function of neutrino energy for $E > 5$ MeV and found it to be of a constant value $\approx 0.3$. SNO has also measured \cite{Aharmim:2011yq} the neutral current interaction rate of solar neutrinos to be consistent with predictions of the standard solar model \cite{Bahcall:2004fg}. These results, together with CHOOZ bound on $\theta_{13}$, provide a 5 $\sigma$ evidence for the matter effects in solar neutrinos \cite{Fogli:2005cq}. The measurements of the ${}^{71}Ga$ experiments \cite{Hampel:1998xg,Abdurashitov:2009tn,Altmann:2005ix} imply that $P_{\rm ee} > 0.5$ for neutrino energies $E < 0.5$ MeV. This increase in $P_{\rm ee}$ at lower energies can be explained only if $\Delta_{21}$ is {\it positive}.

The up going atmospheric neutrinos travel thousands of kilometers through earth, during which they undergo forward elastic scattering with earth matter. They also experience the matter effect \cite{Wolfenstein:1977ue,Wolfenstein:1979ni} which modifies their of survival and oscillation probabilities. As in the case of the solar neutrinos, this modification depends on the sign of the mass-square difference which, in this case, is $\Delta_{31}$. Given the different magnitudes of
\begin{itemize}
\item $\Delta_{21}$ and $\Delta_{31}$, 
\item the energies of solar and atmospheric neutrinos and 
\item the solar and earth matter densities,
\end{itemize}
the matter modification of atmospheric neutrino probabilities are of a different mathematical form compared to their solar neutrino counterparts \cite{Parke:1986jy,Akhmedov:2004ny}. An observation of these matter modifications can establish the sign of $\Delta_{31}$. The case of positive $\Delta_{31}$ is labelled normal hierarchy (NH) and that of negative $\Delta_{31}$ inverted hierarchy (IH). A number of studies~\cite{Petcov:2005rv,Gandhi:2007td,Ghosh:2013mga,Ajmi:2015uda} considered matter modified oscillations and explored the sensitivity of future atmospheric neutrino detectors to determine whether hierarchy is NH or IH. Recently, Super-Kamiokande experiment analyzed their data using the hypothesis of matter modified oscillations. Their results indicate that the vacuum oscillations are disfavoured at $1.6~\sigma$ only \cite{Abe:2017aap}. They prefer NH and disfavour IH at $1.7~\sigma$.

At present, the most precise determination of $|\Delta_{31}|$ and $\sin^2 2 \theta_{23}$ comes from the muon neutrino disappearance data of the accelerator neutrino experiments, such as  MINOS~\cite{Michael:2006rx}, T2K \cite{Abe:2013fuq} and NO$\nu$A \cite{Adamson:2017qqn}. For baselines less than $1000$ km, the matter effects lead to negligibly small changes in $\nu_\mu/ \bar{\nu}_\mu $ survival probabilities \cite{Choubey:2005zy,Gandhi:2004bj}. Thus, the $\nu_\mu/\bar{\nu}_\mu$ disappearance data of accelerator neutrino experiments lead to essentially the same values of $|\Delta_{31}|$ and $\sin^2 2 \theta_{23}$ for the three cases: (a) vacuum oscillations, (b) matter oscillations with NH and (c) matter oscillations with IH. It is important to develop methods to make a distinction between these three hypotheses. Without such a distinction, it will be impossible to measure the CP violating phase $\delta_{\rm CP}$ in neutrino oscillations because matter effects mimic CP violation \cite{Cervera:2000kp, Barger:2001yr}. In the present work, we explore how this distinction can be made with future atmospheric neutrino data. Disentangling the changes induced in the oscillation probabilities by the matter effects and by $\delta_{\rm CP}$ is non-trivial in general. In an atmospheric neutrino detector, the interaction rate of $\nu_\mu$ ($\bar{\nu}_\mu$) depends on both the survival probability $P_{\mu \mu}$ ($P_{\bar{\mu} \bar{\mu}}$) and the oscillation probability $P_{e \mu}$ ($P_{\bar{e} \bar{\mu}}$). However, it has been shown that the sensitivity of this rate to matter effects does not depend on the value of $\delta_{\rm CP}$ \cite{Gandhi:2007td}. Hence, this data can lead to a distinction between the three hypotheses, independent of the value of $\delta_{\rm CP}$. 

The $\nu_{e}/\bar{\nu}_{e}$ appearance data in long-baseline accelerator neutrino experiments is sensitive to matter effects \cite{Lipari:1999wy,Narayan:1999ck}. A precise measurement of the oscillation probabilities, $P_{\mu e}$ and $P_{\bar{\mu} \bar{e}}$, in principle, can make a distinction among the three possibilities. However, this data is also sensitive to $\delta_{\rm CP}$ which at present is poorly determined. A given measured values of $P_{\mu e}$ and $P_{\bar{\mu} \bar{e}}$ can have three solutions \cite{Mena:2004sa, Prakash:2012az}: 
\begin{itemize}
\item vacuum oscillations with $\delta_{\rm CP}^1$,
\item NH matter oscillations with $\delta_{\rm CP}^2$ and
\item IH matter oscillations with $\delta_{\rm CP}^3$.
\end{itemize} 
For each type of oscillating hypothesis, the value of $\delta_{\rm CP}$ determined turns out to be different. Since the determination of $\delta_{\rm CP}$ is one of the important goals of future long-baseline accelerator neutrino experiments, making a distinction between the three hypotheses through independent data is crucial. 

At present, the data of T2K and NO$\nu$A are analyzed using the matter modified oscillation hypothesis with both NH and IH. T2K prefers NH and disfavours IH at 2 $\sigma$ \cite{Abe:2018wpn}. It prefers $\delta_{\rm CP}$ close to $-\pi/2$ for both NH and IH \cite{Abe:2019vii}. NO$\nu$A also prefers NH but it disfavours IH only at 1 $\sigma$ \cite{Acero:2019ksn}. In the case of NH, NO$\nu$A allows the full range of $\delta_{\rm CP}$ within 1 $\sigma$, though it prefers $-\pi/2$ for IH. Very recently, NO$\nu$A updated its results on hierarchy and $\delta_{\rm CP}$ \cite{HimmelNu2020}. It still prefers NH with a wide allowed range of $\delta_{\rm CP}$ but also has a large 1 $\sigma$ allowed region for IH around $\delta_{\rm CP} \simeq -\pi/2$. A combined analysis of the latest data of T2K and NO$\nu$A \cite{Kelly:2020fkv} mildly prefers IH over NH (with a $\Delta \chi^2 = 1.8$) whereas the combined analysis of T2K and NO$\nu$A data along with that of Super-Kamiokande shows a mild preference for NH over IH (with a $\Delta \chi^2 = 2.2$). The combined data of T2K and NO$\nu$A {\bf do not show} any discrimination between vacuum and matter modified oscillations with NH \cite{Bharti:2020gnu}. 
 
\section{Vacuum vs. Matter Modified Oscillations}
The mixing between the neutrino flavor eigenstates and the mass eigenstates is given by
\begin{equation}
\left[
\begin{array}{c}
\nu_{e}\\
\nu_{\mu}\\
\nu_{\tau}
\end{array}\right]=U\left[
\begin{array}{c}
\nu_{1}\\
\nu_{2}\\
\nu_{3}
\end{array}\right],
\end{equation}
where $U$ is a $3 \times 3$ unitary PMNS matrix. It is parameterized as
\begin{equation}
U=\left[
\begin{array}{c c c}
1 & 0&0\\
0&c_{23}&s_{23}\\
0&-s_{23}&c_{23}
\end{array}\right]
\left[
\begin{array}{c c c}
c_{13} & 0&s_{13} e^{-i \delta_{\rm CP}}\\
0&1&0\\
-s_{13} e^{i \delta_{\rm CP}}&0&c_{13}
\end{array}\right]
\left[
\begin{array}{c c c}
c_{12} & s_{12}&0\\
-s_{12}&c_{12}&0\\
0&0&1
\end{array}\right].
\end{equation}
For neutrino propagation in vacuum, the oscillation probabilities depend on the six parameters: the two mass-squared differences, $\Delta_{21}=m_{2}^{2}-m_{1}^{2}$ and $\Delta_{31}=m_{3}^{2}-m_{1}^{2}$, the three mixing angles and $\delta_{\rm CP}$. At present, $\Delta_{21}$, $|\Delta_{31}|$, $\theta_{12}$ and $\theta_{13}$ are measured quite precisely. In case of the third mixing angle, $\sin^{2} 2 \theta_{23}$ is measured to be close to $1$  but $\sin^{2} \theta_{23}$ has a rather large range of $(0.4 - 0.64)$. As mentioned in the introduction, the sign of $\Delta_{31}$ is not known at present. 

The effect of neutrino propagation in matter is parameterized by the Wolfenstein matter term $ A = 0.76 \times 10^{-4} \, \rho ~{\rm(in~gm/cc)} \, E{\rm(in~GeV)}$~\cite{Wolfenstein:1977ue,Wolfenstein:1979ni}. Inclusion of this matter term in neutrino evolution induces a change in the mass-square differences and the mixing angles and hence in the probabilities. In this work, we study how this change can be utilized to make a distinction between vacuum and matter modified oscillations. This change depends on not only the matter term but also on the sign of $\Delta_{31}$. Hence we consider both positive and negative values of $\Delta_{31}$.

\begin{figure}[h]
	\centering
	\begin{minipage}{\textwidth}
		\centering
		\includegraphics[width=.49\linewidth]{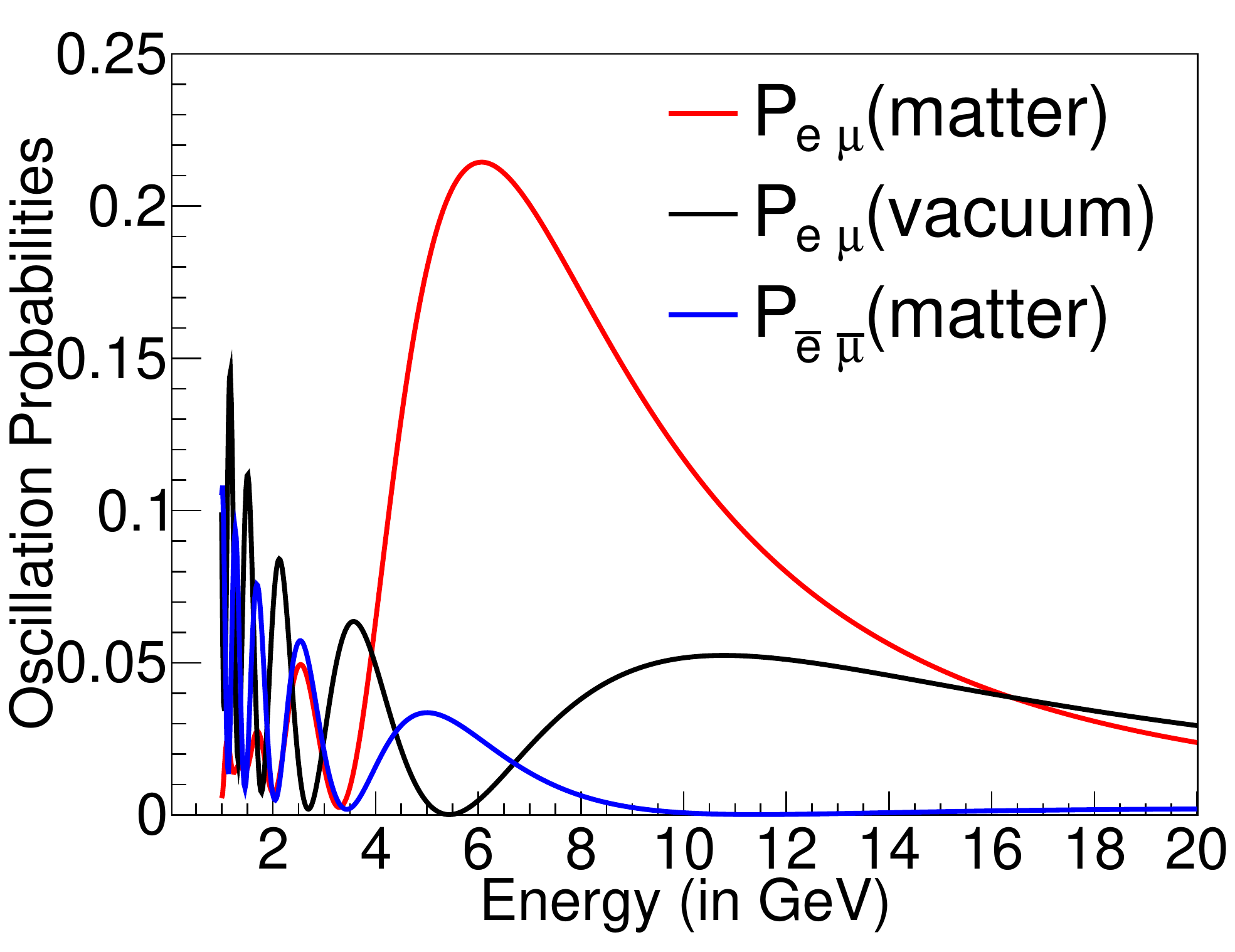}
		\hfill
		\includegraphics[width=.49\linewidth]{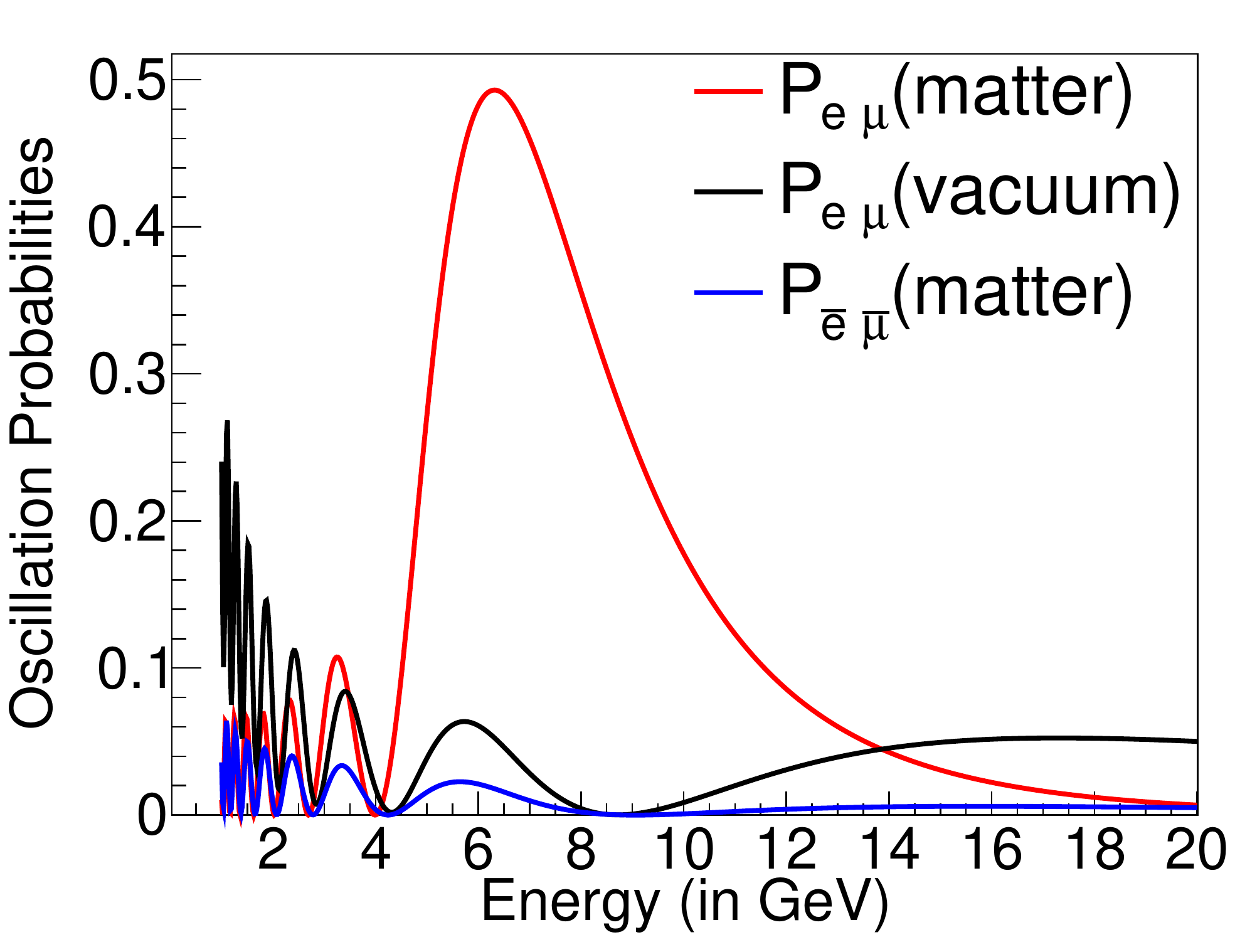}
	\end{minipage}
	\begin{minipage}{\textwidth}
		\centering
		\includegraphics[width=.49\linewidth]{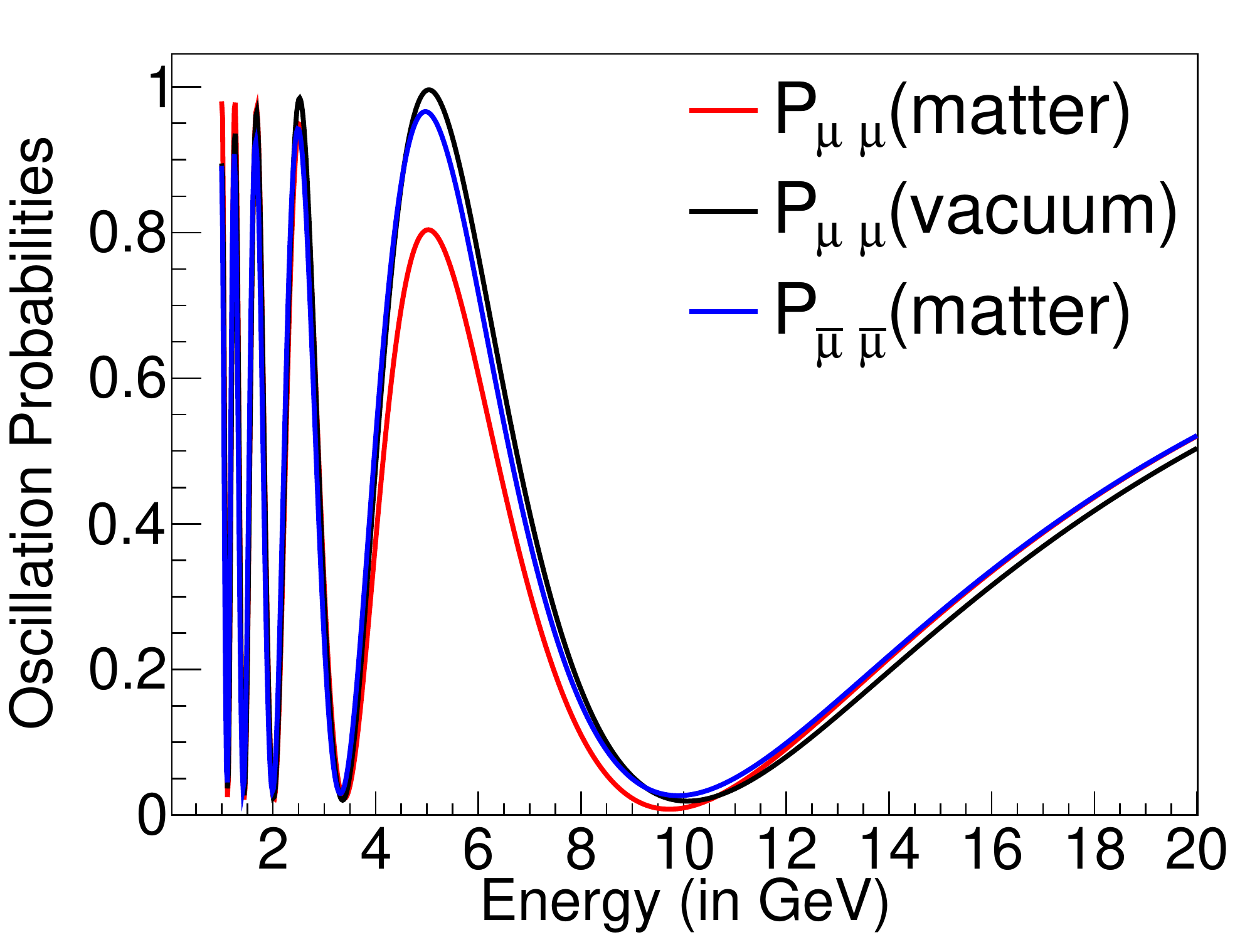}
		\hfill
		\includegraphics[width=.49\linewidth]{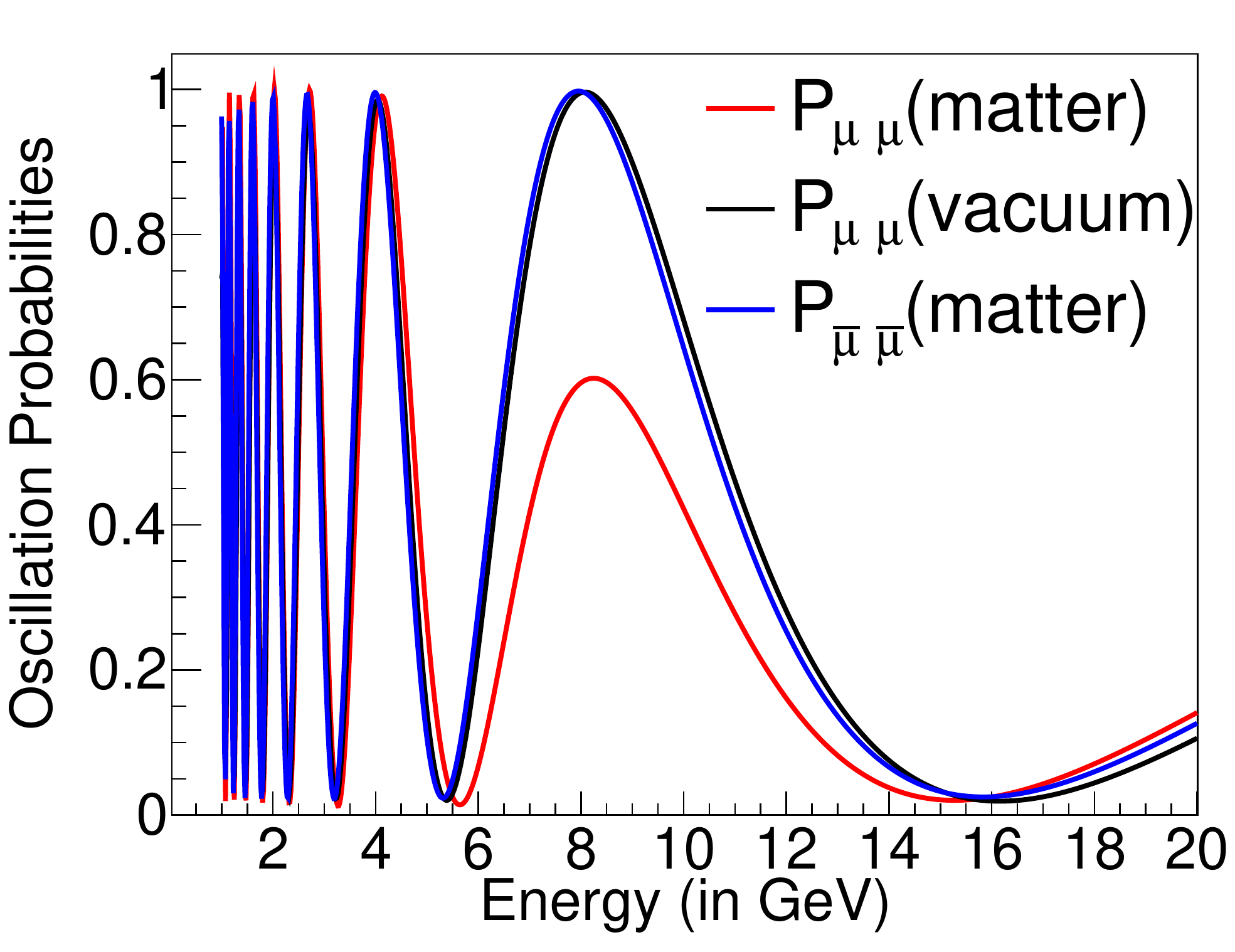}
	\end{minipage}  
	\caption{\label{fig:1}\footnotesize{
Probability vs Energy plots for the case of NH. The top row shows the oscillation probabilities $P_{e \mu}$ and $P_{\bar{e} \bar{\mu}}$ for L = 5000 km (left panel) and L = 8000 km (right panel). The bottom row shows  the survival probabilities $P_{\mu \mu}$ and $P_{\bar{\mu} \bar{\mu}}$ for L = 5000 km (left panel) and L = 8000 km (right panel). The neutrino parameters used for generating these plots, $\sin^2 \theta_{12} = 0.310$, $\sin^2 \theta_{13} = 0.02240$, $\sin^2 \theta_{23} = 0.582$, $\Delta_{21} = 7.39 \times 10^{-5}$ eV$^2$ and $\Delta_{31} = 2.525 \times 10^{-3}$ eV$^2$, are the global best-fit values~\cite{Esteban:2018azc}. The CP violating phase $\delta_{\rm CP}$ is taken to $0$.
 }}
\end{figure}

We first study the difference between matter and vacuum oscillation probabilities for two representative path-lengths for atmospheric neutrinos, $L=5000$ km and $L=8000$ km. Fig:\ref{fig:1} shows the plots of neutrino and anti-neutrino oscillation probabilities $P_{e \mu}$ and $P_{\bar{e} \bar{\mu}}$ and survival probabilities $P_{\mu \mu}$ and $P_{\bar{\mu} \bar{\mu}}$ for vacuum oscillations as well as for matter modified oscillations with NH. The matter modified probabilities are calculated numerically using the code {\bf nuCraft} \cite{Wallraff:2014qka}, which uses the earth density profile of the PREM model \cite{PREM}. From the expressions for the oscillation probabilities, it can be shown that the probabilities for IH can be obtained via the relations $P_{\mu \mu} (IH)= P_{\bar{\mu} \bar{\mu}} (NH)$, $P_{\bar{\mu} \bar{\mu}} (IH) = P_{\mu \mu} (NH)$, $P_{\mu e} (IH) = P_{\bar{\mu} \bar{e}} (NH)$ and $P_{\bar{e} \bar{\mu}} (IH) = P_{e \mu} (NH)$~\cite{Gandhi:2004bj}. From Fig:\ref{fig:1}, we see that, the matter effects increase the peak value of $P_{\mu e}$ from $0.05$ to $0.2$ ($0.5$) for L = 5000 (8000) km and these peak values occur at nearly the same energy where $P_{\mu \mu}$ also peaks. Conservation of probability implies that either $P_{\mu \mu}$ or $P_{\mu \tau}$ should decrease significantly. The maxima of $P_{\mu \mu}$ generally coincide with the minima of $P_{\mu \tau}$ and it can be shown that the change in the value of $P_{\mu \tau}$ near its minimum is very small \cite{Gandhi:2004bj,Gandhi:2004md}. Hence, most of the reduction in the probability occurs in $P_{\mu \mu}$. Thus $P_{\mu \mu}$ for matter oscillations with NH is lower than $P_{\mu \mu}$ for vacuum oscillations over a wide range of energies and path-lengths. But $P_{\bar{\mu} \bar{\mu}}$ is essentially the same for both the cases. In the case of IH, the situation is reversed. Therefore, to study the difference of vacuum oscillations from matter modified oscillations of either sign, it is important to measure neutrino and anti-neutrino event rates separately. In this work, we study the sensitivity of Iron Calorimeter (ICAL) at the India-based Neutrino Observatory (INO) to make a distinction between vacuum and matter modified oscillations using atmospheric neutrino data. The charge identification capability of ICAL leads to a very good sensitivity for this distinction \cite{Kumar:2017sdq}.

\section{Methodology}

The atmospheric neutrinos consist of $\nu_{\mu}$, $\bar{\nu}_{\mu}$, $\nu_{e}$ and $\bar{\nu}_{e}$. The ICAL at INO is a 50 kt magnetized iron calorimeter whose iron plates are interspersed with the active detector elements, Resistive Plate Chambers (RPCs). The charge current (CC) interactions of the neutrinos in the detector produce $\mu^-$ or $\mu^+$ or $e^-$ or $e^+$ depending on the flavor of the initial neutrino.

We have used NUANCE event generator \cite{Casper:2002sd} to simulate the atmospheric neutrino events used in this study. It generates neutrino events using atmospheric neutrino fluxes and the relevant cross sections. For a generated event, NUANCE gives the information on the particle ID and the momenta of all interacting particles. The information of the final state particles is given as an input to a GEANT4 simulator of ICAL. This simulator mimics the response of ICAL and generates the electronic signals of the detector in the form of a hit bank information as the output. A reconstruction program sifts through the hit bank information of each event and tries to reconstruct a track. Electrons and positrons in the final state produce a shower and quickly lose their energy. Identifying such particles and reconstructing their energy is an extremely difficult problem. Muons, being minimum ionizing particles, pass through many layers of iron, leaving behind localized hits in the RPCs. Using this hit information, the track of the muon can be reconstructed. Because of the magnetic field, this track will be curved and the bending of the track is opposite for negative and positive muons. Thus ICAL can distinguish between the CC interactions of $\nu_{\mu}$ and $\bar{\nu}_{\mu}$. If a track is reconstructed, the event is considered to be a CC interaction of $\nu_{\mu} /\bar{\nu}_{\mu}$. The charge, the momentum and the initial direction ($\cos \theta_{\rm track}$), of a reconstructed track, are also calculated from the track properties \cite{Bhattacharya:2014tha}.

We have generated un-oscillated atmospheric neutrino events for 500 years of exposure, using NUANCE. In generating these events, the neutrino fluxes at Kamioka are used as input along with ICAL geometry. The $\nu_{\mu}/\bar{\nu}_{\mu}$ CC events are given as input to GEANT4 and the GEANT4 output is processed by the reconstruction code. Events for which one or more tracks are reconstructed are stored along with the charge, the momentum and the initial direction of the track with the largest momentum. These track variables will be used later in the analysis to bin the events. In the case of $\nu_e/\bar{\nu}_e$ CC events, the electron/positron are redefined to be $\mu^-/\mu^+$ and the events are processed through GEANT4 and the reconstruction code. Once again the charge, the momentum and the initial direction of the track with the largest momentum are stored. This redefinition of $\nu_e/\bar{\nu}_e$ CC events is done so that the events which undergo $\nu_{e}(\bar{\nu}_e) \rightarrow \nu_{\mu}(\bar{\nu}_{\mu})$ oscillation are properly taken into account in our analysis.

We used accept/reject method on the un-oscillated sample to obtain the oscillated event sample. We calculated the vacuum oscillation probabilities, $P_{\mu \mu}$, $P_{\bar{\mu} \bar{\mu}}$, $P_{e \mu}$ and $P_{\bar{e} \bar{\mu}}$, using the formula for three flavor oscillations. The corresponding matter modified probabilities, for both signs of $\Delta_{31}$, are calculated numerically using the code {\bf nuCraft} \cite{Wallraff:2014qka}. The accept/reject method is applied to $\nu_{\mu}(\bar{\nu}_{\mu})$ CC events using $P_{\mu \mu} (P_{\bar{\mu} \bar{\mu}})$ to obtain the muon events due to the survival of $\nu_{\mu}/\bar{\nu}_{\mu}$. The same method is applied to $\nu_{e}(\bar{\nu}_{e})$ CC events using $P_{e \mu} (P_{\bar{e} \bar{\mu}})$ to obtain the muon events due to the oscillation of $\nu_{e}/\bar{\nu}_{e}$. The track information for each of the selected events is taken from the simulation described in the previous paragraph.

\section{Results}
Using the procedure described in the previous section, we generate the muon event sample for matter modified oscillations with $\Delta_{31}$ positive.  In calculating the oscillation probabilities for this case, we used the following values of neutrino parameters as inputs \cite{Esteban:2018azc}: $\sin^2 \theta_{12} = 0.310$, $\sin^2 \theta_{13} = 0.02240$, $\sin^2 \theta_{23} = 0.582$, $\Delta_{31} = 2.525 \times 10^{-3}$ eV$^2$ and $\Delta_{21} = 7.39 \times 10^{-5}$ eV$^2$. We first do our calculation with the input value $\delta_{\rm CP} = 0$. We later show that our sensitivity to matter effects does not depend on the input value of $\delta_{\rm CP}$. The generated sample is divided into $\mu^-$ and $\mu^+$ samples and is further subdivided into $17$ track momentum bins and $90$ track direction bins. The momentum bins are $(1,2)$, $(2,2.2)$, $(2.2,2.4)$, $(2.4,2.6)$, $(2.6,2.8)$, $(2.8,3.0)$, $(3.0,3.5)$, $(3.5,4.0)$, $(4.0,4.5)$, $(4.5,5.0)$, $(5.0,6.0)$, $(6.0,7.5)$, $(7.5,9.0)$, $(9.0,11.0)$, $(11.0,14.0)$, $(14.0,20.0)$, $(20.0,100.0)$. We considered only those events with track momentum greater than $1$ GeV because such events lead to good track reconstruction. The signature of oscillations is very small for the down going events and it is almost impossible to reconstruct tracks of muons moving in horizontal direction. Therefore, we considered only the up going events with $\cos \theta_{\rm track}$ in the range $(0.1,1)$. Since ICAL can reconstruct the muon direction very accurately, we have subdivided the above range into bins of equal width $0.01$. Using this procedure, we have two binned event samples, $N^{data,\mu^-}_{ij}$ and $N^{data,\mu^+}_{ij}$, which we treat as data. Here $i$ refers to the track momentum bin and varies from $1$ to $17$ and $j$ refers to $\cos \theta_{\rm track}$ bin and varies from $1$ to $90$. 

\begin{figure}[h]
	\centering
	\begin{minipage}{\textwidth}
		\centering
		\includegraphics[width=\linewidth]{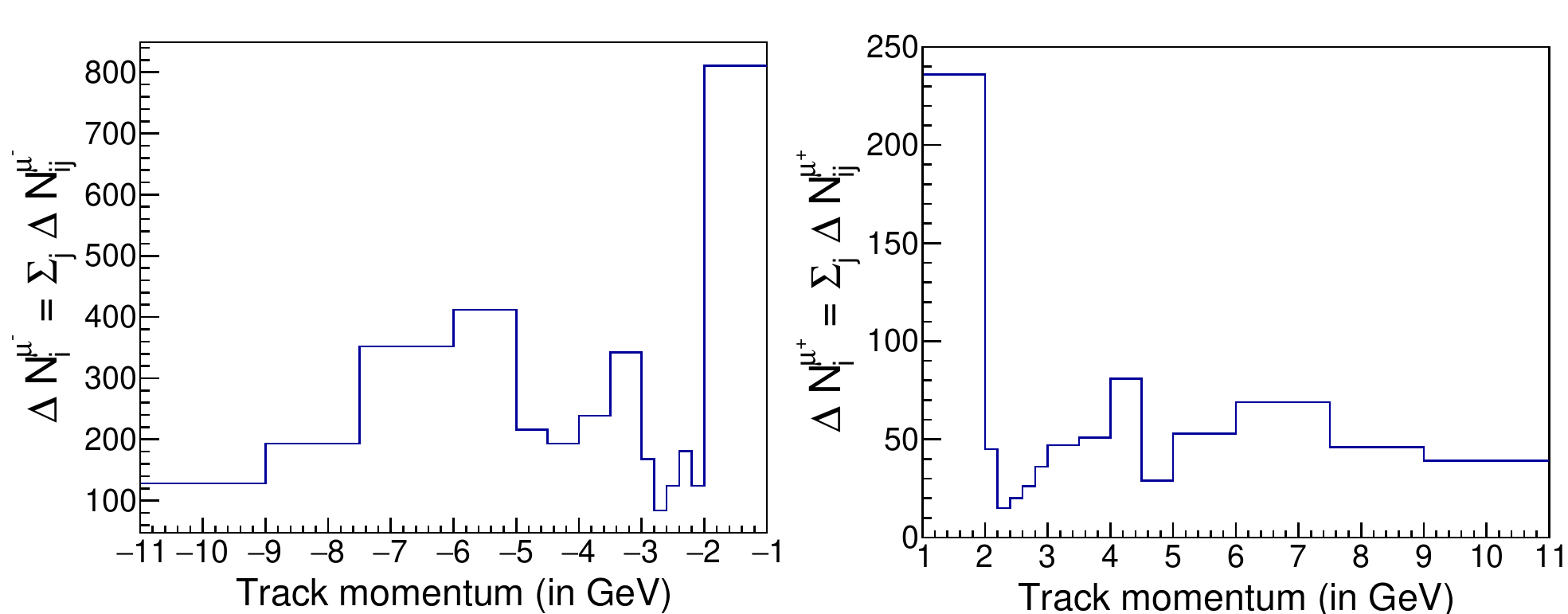}
	\end{minipage}
	\begin{minipage}{\textwidth}
		\centering
		\includegraphics[width=\linewidth]{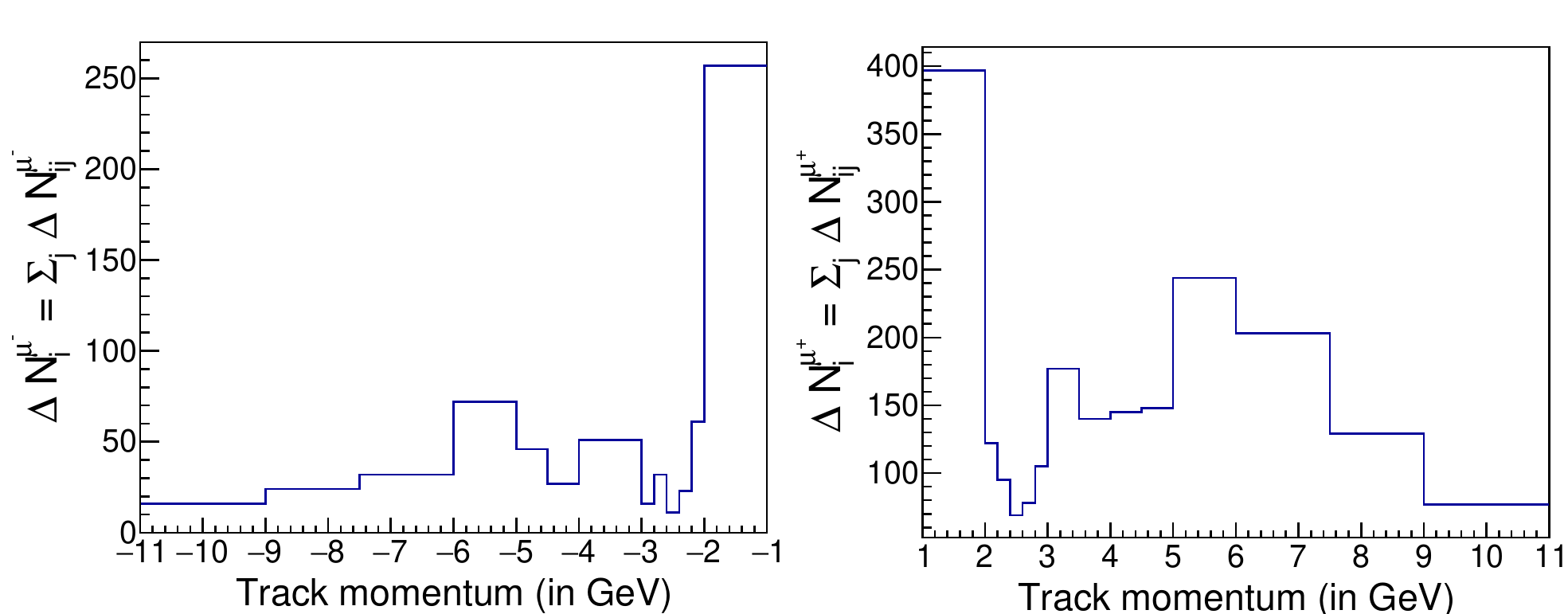}
	\end{minipage} 
	\caption{\label{fig:3}\footnotesize{
The difference between the number of muon events for matter vs. vacuum oscillations ($\Delta N^{\mu^\mp} _i$) as a function of track momentum. The plots in the left (right) panels are for $\mu^- (\mu^+)$ events. The plots in the top (bottom) panels are for $\Delta_{31}$ positive (negative).
}}
\end{figure}

\begin{figure}[h]
	\centering
	\begin{minipage}{\textwidth}
		\centering
		\includegraphics[width=\linewidth]{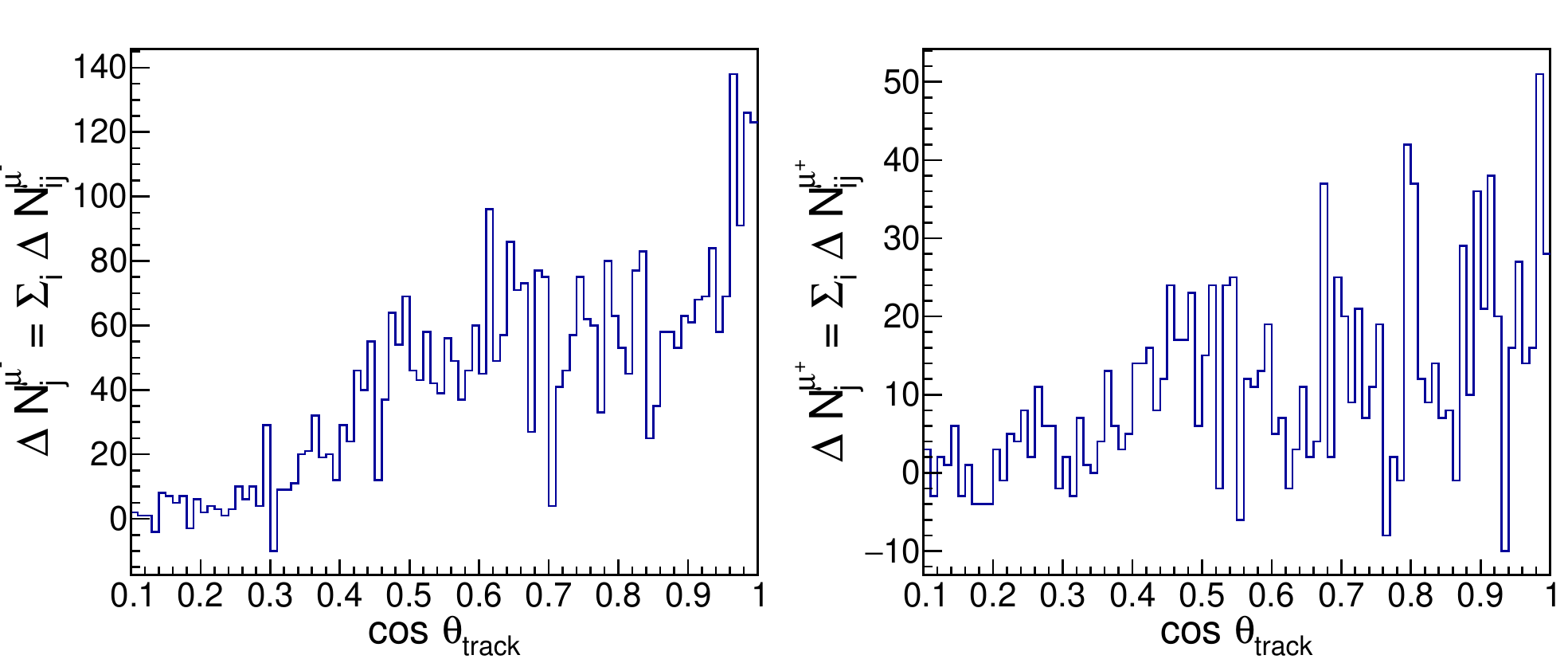}
	\end{minipage}
	\begin{minipage}{\textwidth}
		\centering
		\includegraphics[width=\linewidth]{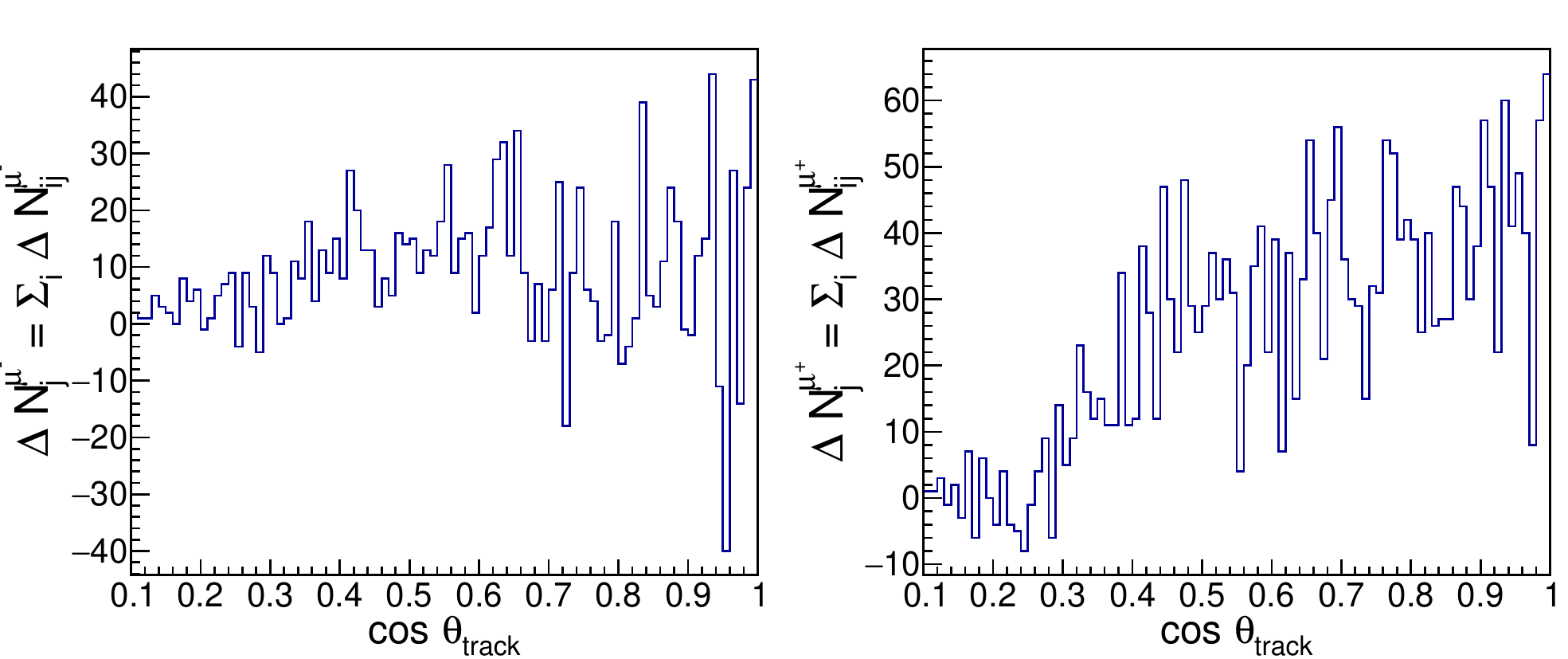}
	\end{minipage}
	\caption{\label{fig:4}\footnotesize{
The difference between the number of muon events for matter vs. vacuum oscillations ($\Delta N^{\mu^\mp} _j$) as a function of $\cos \theta_{\rm track}$. The plots in the left (right) panels are for $\mu^- (\mu^+)$ events. The plots in the top (bottom) panels are for $\Delta_{31}$ positive (negative).
}}
\end{figure}

We consider the vacuum oscillations as a hypothesis to be tested against the data samples described above. Using the vacuum oscillation hypothesis, two other event samples, $N^{vac,\mu^-}_{ij}$ and $N^{vac,\mu^+}_{ij}$, are generated using the same procedure described in the previous paragraph. In calculating the vacuum oscillation probabilities the five inputs, $\sin^2 \theta_{12} = 0.310$, $\sin^2 \theta_{13} = 0.02240$, $\Delta_{31} = 2.525 \times 10^{-3}$ eV$^2$, $\Delta_{21} = 7.39 \times 10^{-5}$ eV$^2$ and $\delta_{\rm CP}= 0$, are held fixed. The test values of $\sin^2 \theta_{23}$ are varied in the range $(0.4,0.64)$. To quantify the difference between matter and vacuum oscillations, we define $\Delta N_{ij}^{\mu^-} = N^{vac,\mu^-}_{ij} - N^{data,\mu^-}_{ij}$ and $\Delta N_{ij}^{\mu^+} = N^{vac,\mu^+}_{ij} - N^{data,\mu^+}_{ij}$. In Fig:\ref{fig:3}, we plot $\Delta N^{\mu^-} _i = \Sigma_j \Delta N_{ij}^{\mu^-}$ and $\Delta N^{\mu^+} _i = \Sigma_j \Delta N_{ij}^{\mu^+}$ as a function of track momentum. Fig:\ref{fig:4} gives the plots of $\Delta N^{\mu^-} _j = \Sigma_i \Delta N_{ij}^{\mu^-}$ and $\Delta N^{\mu^+} _j = \Sigma_i \Delta N_{ij}^{\mu^+}$ as a function of track direction. We see that $\Delta N_i^{\mu^\mp}$, in general, are positive because the matter effects suppress the peak values of $P_{\mu \mu}$. We also see that the magnitude of $\Delta N_i^{\mu^-}$ is larger than that of $\Delta N_i^{\mu^+}$ for $\Delta_{31}$ positive. The situation is reversed when $\Delta_{31}$ is negative. This, of course, is a reflection of the fact that the matter effects are more pronounced in $P_{\mu \mu}$ for positive value of $\Delta_{31}$ and in $P_{\bar{\mu} \bar{\mu}}$ for negative values of $\Delta_{31}$. A similar pattern is also seen in $\Delta N_j^{\mu^\mp}$, for the same reasons.   

We calculate the test event samples $N^{test,\mu^-}_{ij}$ and $N^{test,\mu^+}_{ij}$ as
\begin{equation}
  N_{ij}^{test,\mu^-/\mu^+}=N_{ij}^{vac,\mu^-/\mu^+}[1+\pi^k_{ij} \xi_k],
\end{equation}
where we introduced three systematic errors $\pi^k_{ij}$ $(k = 1,2,3)$, each with its pull parameter $\xi_k$. The first of these is the systematic error in flux normalization which is independent of track momentum and track direction. The second systematic error depends on the track momentum and the  third one depends on track direction. The method of calculation of the last two systematic errors is described in the appendix. We computed the $\chi^2$ between the data and the test event samples by
\begin{equation}
\chi^2=\chi^2(\mu^-)+\chi^2(\mu^+)+\xi_k^2,
\end{equation}
where
\begin{eqnarray}
\chi^2(\mu^-)=\Sigma_{i=1}^{17} \Sigma_{j=1}^{90} 2\left[\left(N_{ij}^{test,\mu^-} - N_{ij}^{data,\mu^-}\right) - N_{ij}^{data,\mu^-} \ln\left(\frac{N_{ij}^{test,\mu^-}}{N_{ij}^{data,\mu^-}}\right)\right],\\
\chi^2(\mu^+)=\Sigma_{i=1}^{17} \Sigma_{j=1}^{90} 2\left[\left(N_{ij}^{test,\mu^+} - N_{ij}^{data,\mu^+}\right) - N_{ij}^{data,\mu^+} \ln\left(\frac{N_{ij}^{test,\mu^+}}{N_{ij}^{data,\mu^+}}\right)\right], 
\end{eqnarray}
and priors on the pull parameters $\xi_k^2$ are added. For each test value of $\sin^2\theta_{23}$, the minimum value of $\chi^2$ is computed by varying the pull parameters $\xi_k$ in the range $(-3,3)$ in steps of $0.1$. We obtain the $\chi^2$ for a ten year exposure by dividing the minimum $\chi^2$ by $50$. This $\chi^2$ is a measure of ICAL sensitivity to distinguish vacuum oscillations from matter oscillations with $\Delta_{31}$ positive.

In calculating the vacuum oscillation probabilities, we held $\Delta_{21}, \Delta_{31}, \theta_{12}, \theta_{13}$ and $\delta_{\rm CP}$ fixed. However, each of these parameters has an associated uncertainty. Varying $\Delta_{21}$ and $\theta_{12}$ has very little effect on the atmospheric neutrino oscillation probabilities. So these parameters are kept fixed for the whole calculation. 
The variation in the two parameters, $\Delta_{31}$ and $\sin^2 \theta_{13}$, can lead to a noticeable change in the probabilities and hence in the event rates. Therefore, we varied these two parameters and marginalized over them. The vacuum probabilities are calculated for various different test values of $\sin^2 \theta_{13}$
and $\Delta_{31}$. These test values are chosen within the $\pm 2~\sigma$ ranges of the central values of the respective parameters. The $\chi^2$ is computed with the addition of the following priors
\begin{eqnarray}
\chi^2_{\rm prior}(\theta_{13}) & = & \left(\frac{\sin^2 \theta_{13}^{test} - 0.0224}{\sigma_{\sin^2 \theta_{13}}}\right)^2, \\
\chi^2_{\rm prior}(\Delta_{31}) & = & \left(\frac{\Delta_{31}^{test} - 2.525 \times 10^{-3}}{\sigma_{\Delta_{31}}}\right)^2,
\end{eqnarray}
where $\sigma_{\sin^2\theta_{13}} = 0.00066$ and $\sigma_{\Delta_{31}} = 0.033 \times 10^{-3}$ eV$^2$~\cite{Esteban:2018azc}. We found that the minimum $\chi^2$ occurred when the values of $\sin^2 \theta_{13}$ and $\Delta_{31}$ in the vacuum probability  are the same as the central values.

\begin{figure}[h]
	\centering
	\includegraphics[width=\linewidth]{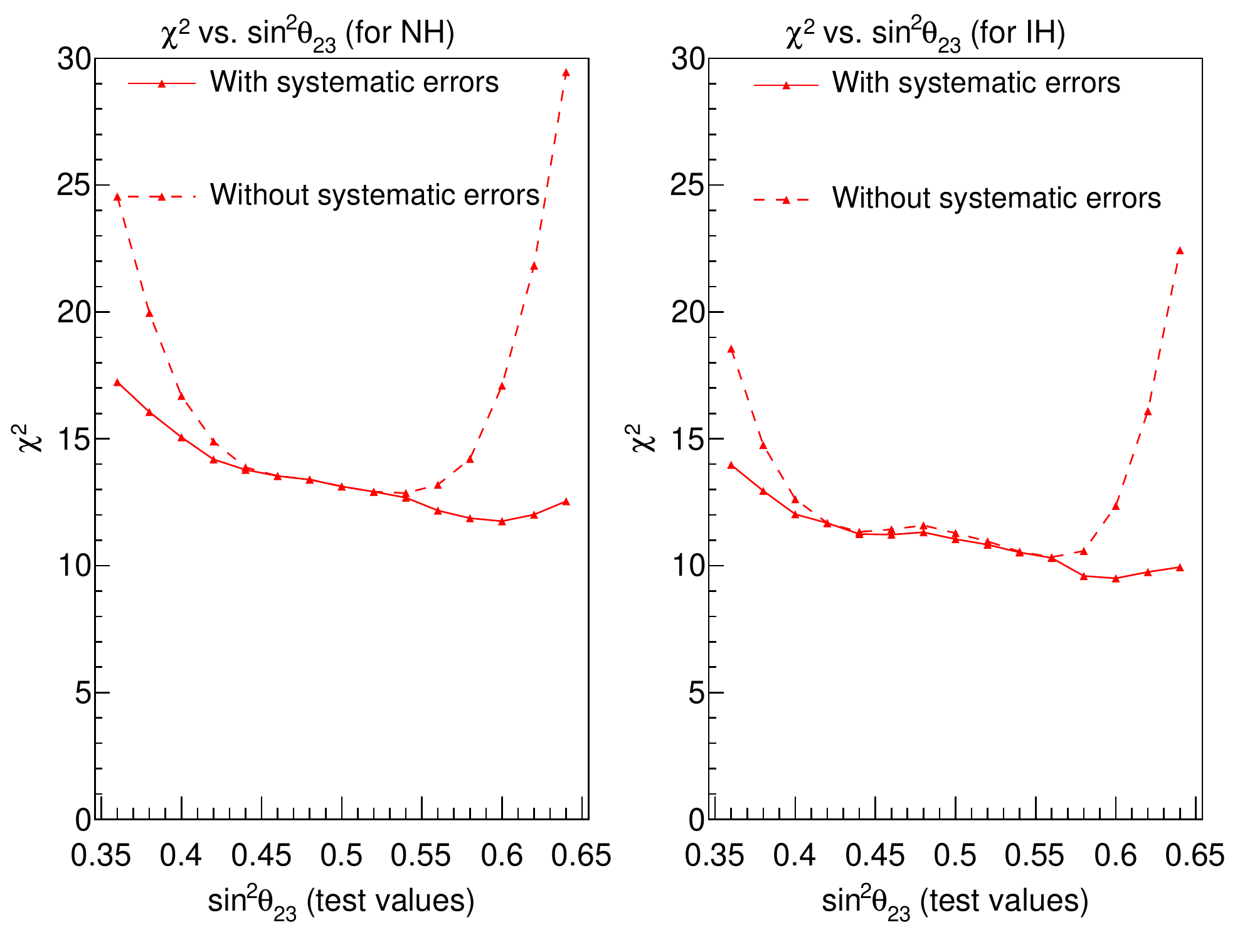}
	\caption{\label{fig:5} \footnotesize{
Sensitivity of ICAL to matter vs. vacuum oscillations as a function of $\sin^2 \theta_{23}$(test). 
The CP violating phase $\delta_{\rm CP}$ is set equal to $0$ for both matter and vacuum oscillations. 
}}
\end{figure}

\begin{figure}[h]
	\centering
	\includegraphics[width=0.8\linewidth]{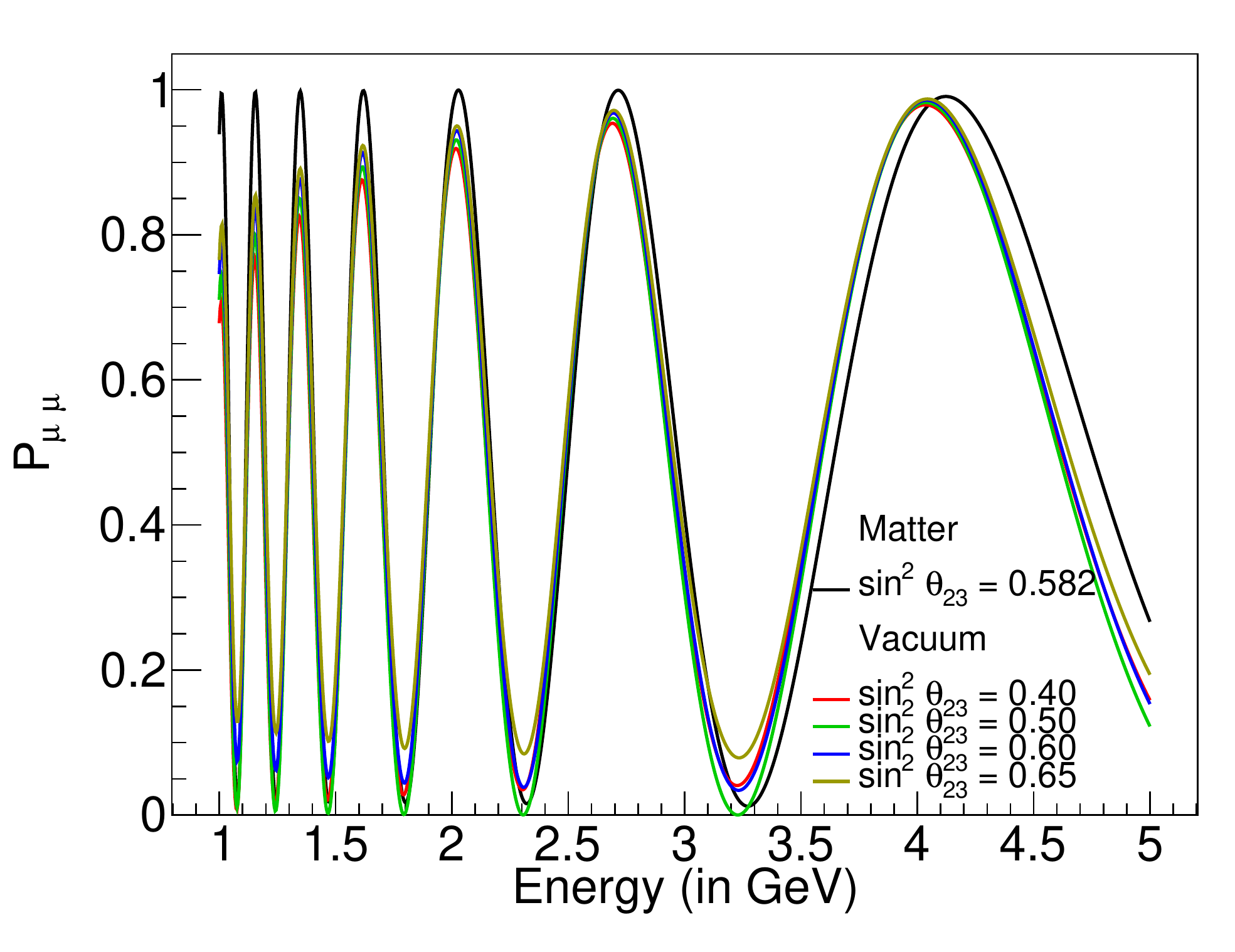}
	\caption{\label{fig:2}\footnotesize{
 $P_{\mu \mu}$ plots for the energy range $E = (1.0, 5.0)$ GeV. Shown are the matter modified
probability for $\sin^2 \theta_{23} = 0.582$ and vacuum probabilities for the test values of $\sin^2 \theta_{23} =
0.4, 0.5, 0.6$ and $0.65$.
}}
\end{figure}

After calculating ICAL sensitivity to distinguish vacuum oscillations from matter oscillations with $\Delta_{31}$ positive, we repeat this calculation for the case where $\Delta_{31}$ is negative.  Our results are shown in Fig:\ref{fig:5} where the left panel is for $\Delta_{31}$ positive and the right panel is for $\Delta_{31}$ negative. Each panel shows the variation of $\chi^2$ as a function of the test values of $\sin^2 \theta_{23}$, without and with the systematic errors.
We see that the $\chi^2$ without systematic errors is relatively flat for $\sin^2\theta_{23}$(test) in the range $(0.4,0.6)$ but rises sharply outside this range. This behaviour can be explained by close examination of the difference between the values of $P_{\mu \mu}$ near its minima for the two cases of matter modified oscillations and vacuum oscillations, as shown in Fig:~\ref{fig:2}. From this figure, we note that, values of $P_{\mu \mu}$ near the low energy minima are very close for matter modified oscillations with NH and for vacuum oscillations with test values of $\sin^2\theta_{23}$ varying in the range $(0.4,0.6)$. But, for test value of $\sin^2 \theta_{23}$ away from this range, these differences become bigger. And the contribution from all these energy values leads to a significant rise in the $\chi^2$. The uncertainty in flux normalization, which is common to all the bins, is fairly large. Due to the systematic error in the flux normalization, the $\chi^2$ occurring due to the differences near the $P_{\mu \mu}$ minima regions is drastically reduced whereas the $\chi^2$ occurring due to the large difference near the $P_{\mu \mu}$ maxima regions is relatively unaffected. Hence, the $\chi^2$ with systematic errors is relatively flat with respect to all the test values of $\sin^2 \theta_{23}$.

We see that $\chi^2_{\rm min} = 11.8$ for $\Delta_{31}$ positive and is $9.5$ for $\Delta_{31}$ negative. Hence, ICAL can rule out vacuum oscillations at better than $3~\sigma$ confidence level, if the matter effects, as prescribed by Wolfenstein \cite{Wolfenstein:1977ue,Wolfenstein:1979ni}, are present. This sensitivity is there for both the signs of $\Delta_{31}$. 
{\color{red} The sensitivity 
obtained is for the input values of $\sin^2 \theta_{23} = 0.582$ and $\sin^2 2 \theta_{13} \approx 0.09$, which are the current
best-fit values from the analysis of global data \cite{Esteban:2018azc}. This sensitivity is slightly larger than that obtained 
for hierarchy discrimination for ICAL in previous studies. For example, fig~5.7 of ref.~\cite{Kumar:2017sdq}, gives the hierarchy
sensitivity of ICAL for different input values of mixing angles and exposures. The ten-year hierarchy sensitivity, 
for $\sin^2 \theta_{23} = 0.6$ and $\sin^2 2 \theta_{13} = 0.1$, is shown to be 11.5, independent of whether
the true hierarchy is normal or inverted. 
}
It should, however, be noted that the method of analysis used here is very different from that used in previous studies. The kinematic variables used here are track momentum and track direction which are obtained from the reconstruction of the GEANT simulation of atmospheric neutrino events \cite{Bhattacharya:2014tha}, whereas in the previous studies the kinematic variables are the NUANCE output information on muon, smeared with resolution functions. A detailed comparison of the results from the old method and from the new method is beyond the scope of this paper.

\begin{figure}[h]
	\centering
	\includegraphics[width=\linewidth]{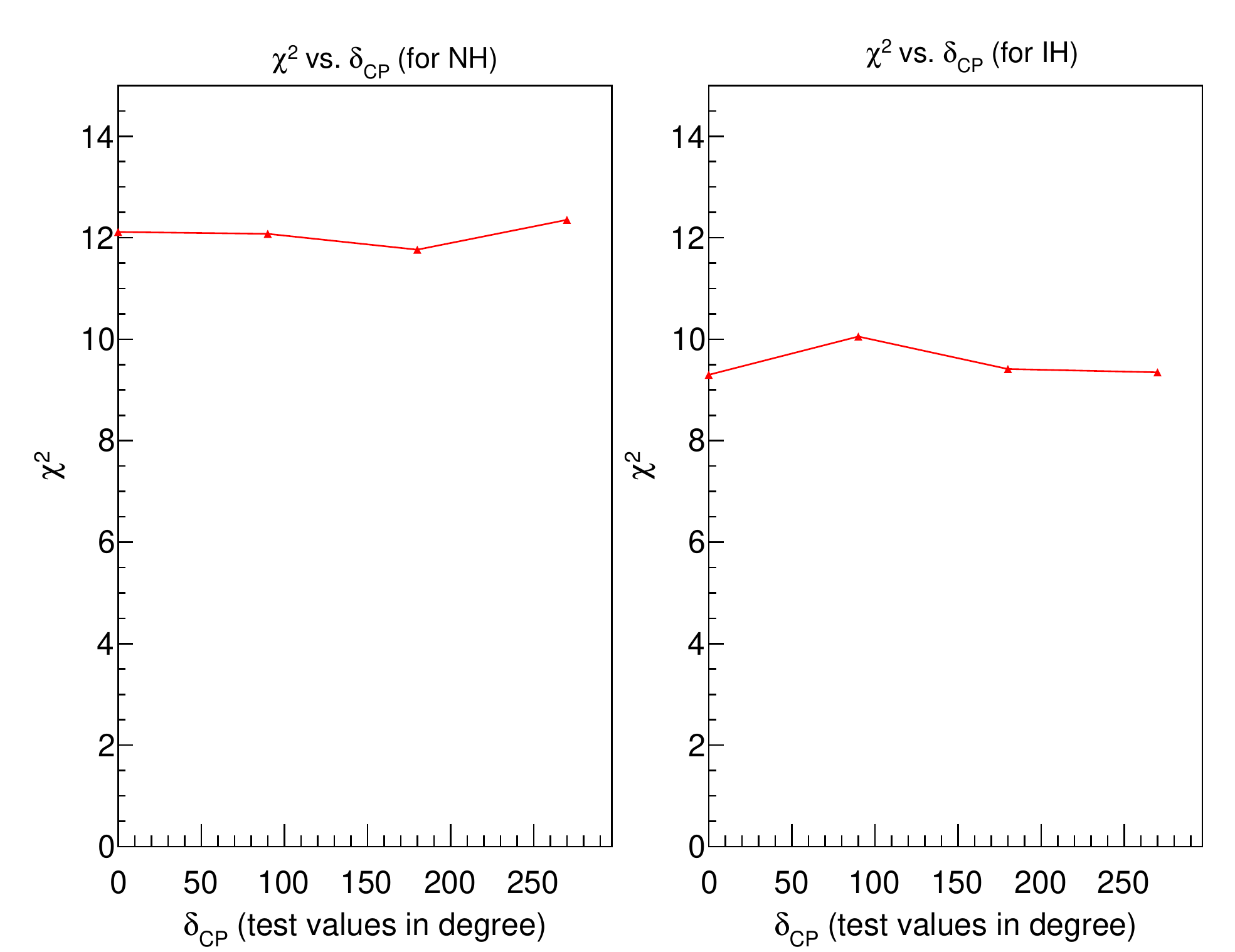}
	\caption{\label{fig:6} \footnotesize{
Sensitivity of ICAL to matter vs. vacuum oscillations as a function of test values of $\delta_{\rm CP}$. This phase is set equal to $0$ for matter oscillations and is varied over four test values for vacuum oscillations.
}}
\end{figure}

The results presented in Fig:\ref{fig:5} assumed $\delta_{\rm CP}$ to be zero for both matter and vacuum oscillations. However we should check the sensitivity if the test value of $\delta_{\rm CP}$ is varied over its full range $(0,360^\circ)$. We performed this calculation where we kept the true value of $\delta_{\rm CP}= 0$ for matter oscillation probabilities and considered the four test values $\delta_{\rm CP}= 0, 90^\circ, 180^\circ, 270^\circ$ for vacuum oscillations. The minimum $\chi^2$, as a function of test $\delta_{\rm CP}$ is shown in Fig:\ref{fig:6}. We see that this marginalization over $\delta_{\rm CP}$ has essentially no effect on the minimum $\chi^2$. The $\chi^2_{\rm min}$ values are $11.7$ for positive $\Delta_{31}$ and $9.5$ for negative $\Delta_{31}$. Recent global fits to neutrino oscillation data yield the best fit value of $\delta_{\rm CP}\approx 270^\circ$ for both signs of $\Delta_{31}$ \cite{Esteban:2018azc}. We have redone our calculations with $\delta_{\rm CP} = 270^\circ$ as our input value in matter oscillation probabilities and marginalized over the full range of the test values of $\delta_{\rm CP}$. The minimum $\chi^2$ occurred for the test value $180^\circ$. The values of minimum $\chi^2$ are $11.8$ for $\Delta_{31}$ positive and $9.3$ for $\Delta_{31}$ negative. 


\begin{figure}[h]
	\centering
	\includegraphics[width=\linewidth]{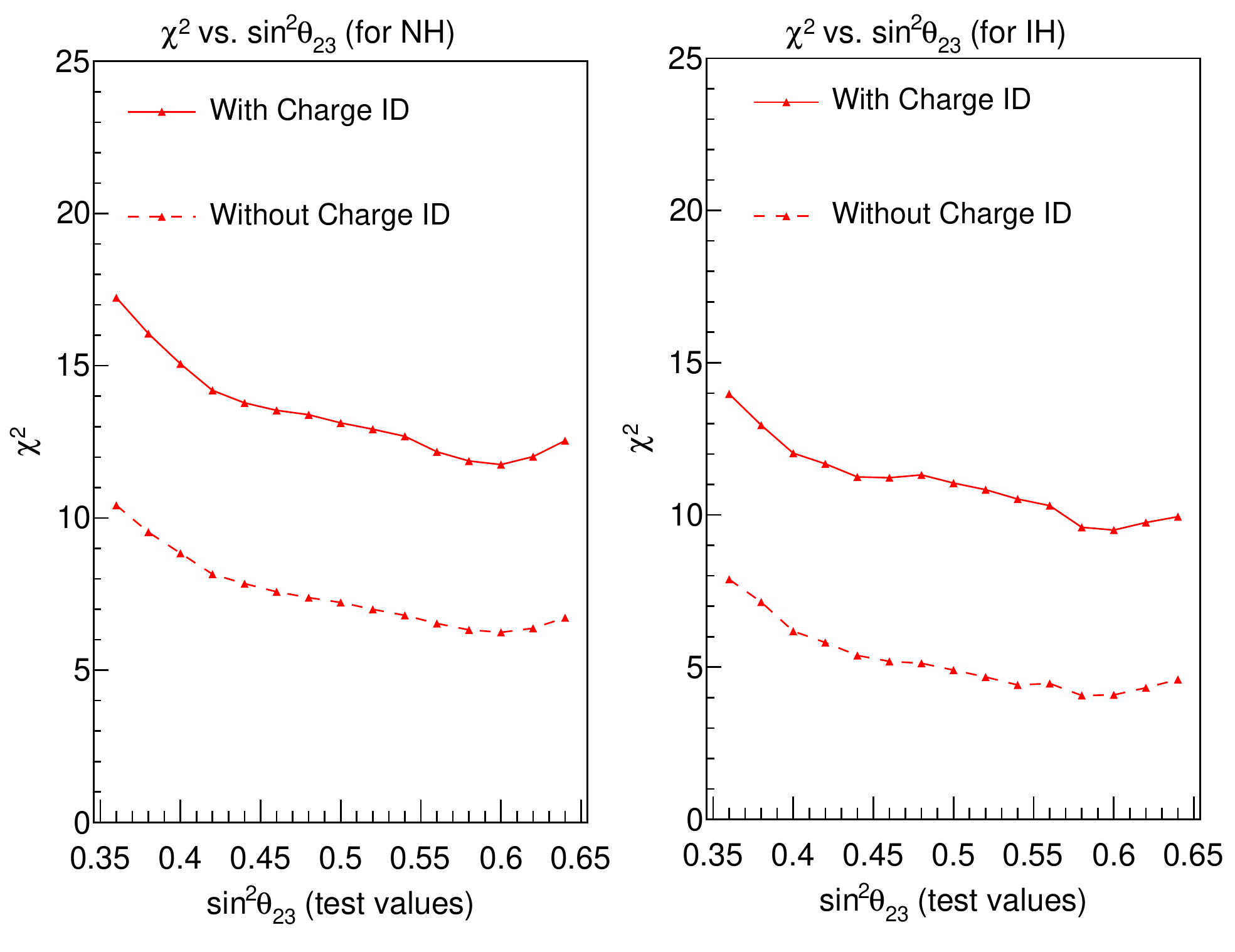}
	\caption{\label{fig:7} \footnotesize{
 Sensitivity of ICAL to matter vs. vacuum oscillations, as a function of test values of 
$\sin^2 \theta_{23}$ \bf{with and without charge identification}.
}}
\end{figure}

It is worth exploring the role of charge identification capability of ICAL in the discrimination sensitivity. To do this, we combined the $\mu^-$ and $\mu^+$ event samples into a single sample and computed the $\chi^2$ between the matter and vacuum oscillated distributions, including the systematic uncertainties and marginalization over the oscillation parameters. The results are shown in Fig:\ref{fig:7}, which show that the sensitivity reduces a factor of $2$ if the charge identification is not there.

\begin{figure}[h]
	\centering
	\includegraphics[width=\linewidth]{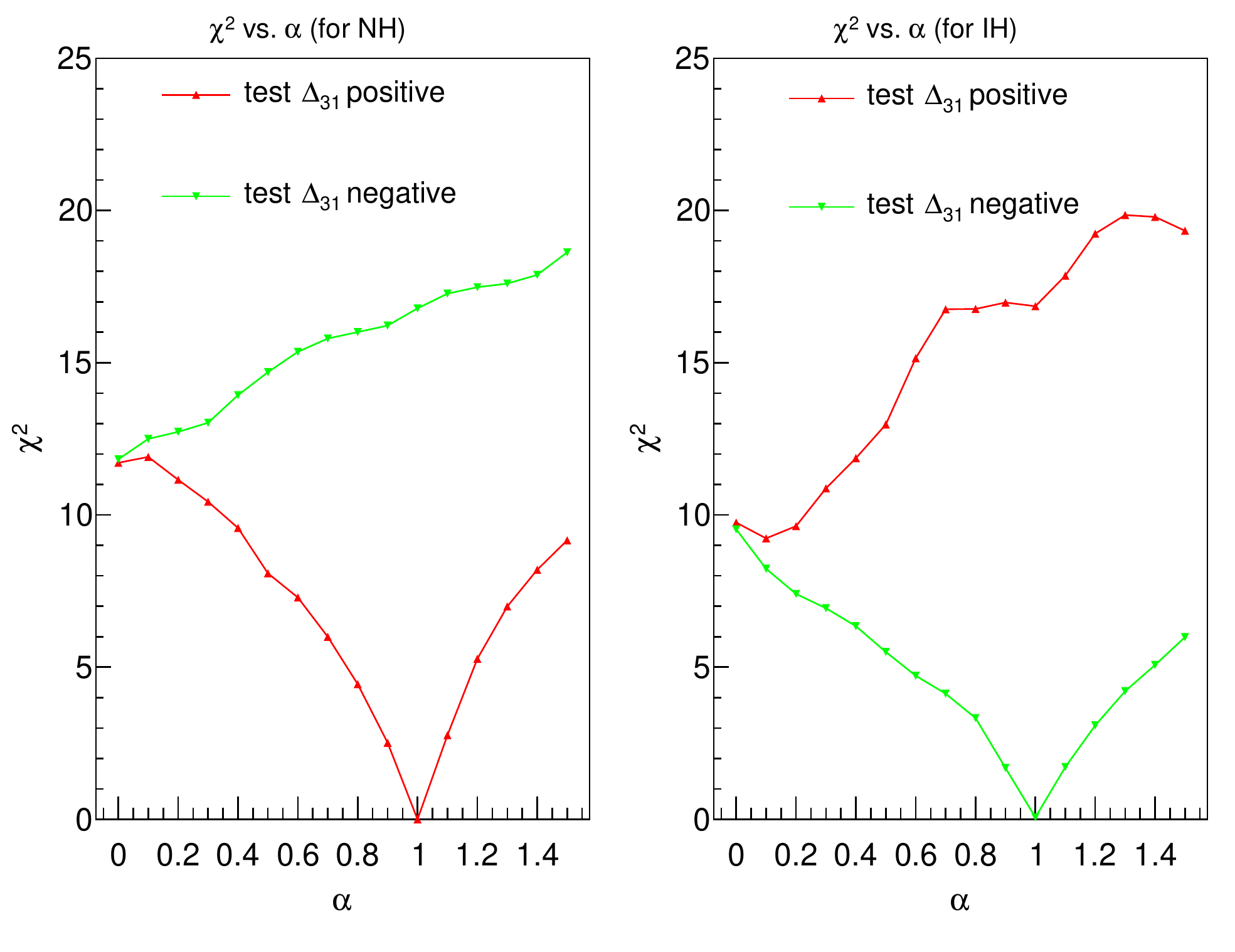}
	\caption{\label{fig:8} \footnotesize{
Sensitivity of ICAL to fractional matter effects for NH (left panel) and IH (right panel).
}}
\end{figure}

Recently Super-Kamiokande collaboration performed an analysis of atmospheric neutrino oscillation data along with external constraints \cite{Abe:2017aap}. They also searched for evidence of matter effects in this data. They parameterized the matter term in the form {\bf $\alpha$ * the standard matter effect} and varied $\alpha$ in the range $(0, 2)$. Vacuum oscillations correspond to $\alpha = 0$ and standard matter oscillations correspond to $\alpha=1$. The best fit point occurred for $\Delta_{31}$ positive and $\alpha=1$. Vacuum oscillations were disfavored with $\Delta \ \chi^2 =5$. Negative $\Delta_{31}$, for all values of $\alpha$, was disfavored with $\Delta \ \chi^2$ in the range $(5 - 6)$. We did a similar analysis for ICAL also. 
We generated the matter modified oscillation events or both NH and IH and tested this "data" against the hypothesis of {\it partial matter effect} oscillations, as parameterized by Super-Kamiokande. Our results are presented in the Fig: \ref{fig:8}, which confirms that the vacuum oscillations, given by $\alpha = 0$, are ruled out with $\chi^2 = 11.5$ for NH (left panel) and with $\chi^2 = 9.5$ for IH (right panel). We also see that ICAL is very effective in ruling out the wrong hierarchy for all values of $\alpha$.

\section{Conclusions}
Both atmospheric neutrino data and accelerator neutrino data give a consistent picture of neutrino oscillations driven by a mass-squared difference of about $2.5 \times 10^{-3}$ eV$^2$ and a nearly maximal mixing angle. However, the present data can not effectively distinguish between vacuum oscillations and matter modified oscillations. It is important to make this distinction because our ability to determine the CP violating phase $\delta_{\rm CP}$ depends on it. In this paper, we have considered the sensitivity of ICAL at INO to make a distinction between these two types of oscillations. We find that a ten year exposure leads to a better than $3~\sigma$ distinction, whether $\Delta_{31}$ is positive or negative. The difference between the matter and vacuum oscillations is significant for neutrinos if $\Delta_{31}$ is positive and for anti-neutrinos if $\Delta_{31}$ is negative. Hence, the charge identification capability of ICAL has an important role in giving rise to such good distinguishing ability. This ability is independent of the true value of $\delta_{\rm CP}$. With no charge identification, the discrimination ability is reduced by half.


\section{appendix}

In the appendix, we describe the method by which we computed the two systematic errors, 
one of which depends on track momentum and the other on track direction. Usually, the
atmospheric neutrino fluxes are listed with three systematic errors: 
\begin{itemize}
\item
Normalization error (common for all the bins)
\item
Tilt error (which depends on neutrino energy)
\item
Direction error (which depends on neutrino direction).
\end{itemize}
The normalization error on neutrino flux translates directly into the normalization of the neutrino event rates in all bins and hence is independent of track momentum and track direction also. The tilt error is taken to be $\pi_{ij}^{\rm tilt} = \pi_i^{\rm tilt} = 0.05*ln(E_\nu/2)$ and the direction error is assumed to be $\pi_{ij}^{\rm nudir} = \pi_j^{\rm nudir} = 0.05*|\cos(\theta_\nu)|$ \cite{GonzalezGarcia:2004wg}, where $E_\nu$ and $\theta_\nu$ are the energy and the zenith angle of the neutrino respectively.

We outline the procedure for the construction of the transfer matrices which convert the systematic errors in neutrino energy and direction to systematic errors in track momentum and direction. To keep the calculations tractable, we make two simplifying assumptions. We assume that track momentum dependent systematic error depends only on neutrino energy dependent systematic error and not on neutrino direction dependent systematic error. In computing this error, we consider only the modulus of the track momentum and do not make a distinction between the charges. Similarly we will also assume that track direction dependent systematic error depends only on the neutrino direction dependent systematic error and not on the neutrino energy dependent systematic error. Here also, we consider only modulus of the cosine of the zenith angle and do not make a distinction between up-going and down-going particles. 

From the simulation of 500 years of un-oscillated events, the $\nu_\mu$-CC and $\bar{\nu}_\mu$-CC events with a well reconstructed track are chosen. These events are classified into ten bins of $E_\nu$ and the event numbers are labeled $N_1$, $N_2$, ... $N_{10}$. The events in the set $N_1$ are further classified into ten bins based on the modulus of track momentum. These event numbers are called $A_{11}$, $A_{21}$, ... $A_{10,1}$, each of which is divided by $N_1$ to get a set of ten fractions $a_{11}$, $a_{21}$, ... $a_{10,1}$. These ten fractions form the first column of the transfer matrix $a$. We repeat this procedure for each of $N_2$, $N_3$, ..., $N_{10}$ to construct the rest of the columns of the $10 \times 10$ matrix $a$. Now we can write a matrix equation of the form
$$
\left[\begin{array}{c} T_1 \\ T_2 \\ . \\ . \\ T_{10} \end{array} \right] = 
\left[\begin{array}{ccccc} a_{11} & a_{12} & . & . & a_{1,10} \\
a_{21} & a_{22} & . & . & a_{2,10} \\
. & . & . & . & . \\
. & . & . & . & . \\
a_{10,1} & a_{10,2} & . & . & a_{10,10} \end{array} \right]
\left[\begin{array}{c} N_1 \\ N_2 \\ . \\ . \\ N_{10} \end{array} \right].
$$ 
The numbers in the column matrix on the LHS, $T_1$, $T_2$, ..., $T_{10}$ are the number of events distributed into ten bins in the modulus of track momentum. Thus, we construct the transfer matrix which  converts the distribution of events in neutrino energy into distribution of events in the magnitude of the track momentum.

We assume that the systematic errors in $N_1$ ($\pi_1^{\rm tilt})$ is independent of the systematic error in $N_2$ ($\pi_2^{\rm tilt})$ and all other systematic errors in other $N_j$s. Therefore the systematic error in $T_1$ ($\pi_1^{\rm trkmm}$) will be
$$
\pi_1^{\rm trkmm} = \sqrt{(a_{11}*\pi_1^{\rm tilt})^2 + (a_{12}*\pi_2^{\rm tilt})^2 + ...
+ (a_{1,10}*\pi_{10}^{\rm tilt})^2}.
$$
Similarly the systematic error in $T_2$ will be
$$
\pi_2^{\rm trkmm} = \sqrt{(a_{21}*\pi_1^{\rm tilt})^2 + (a_{22}*\pi_2^{\rm tilt})^2 + ...
  + (a_{2,10}*\pi_{10}^{\rm tilt})^2}.
$$
This way we can compute all the ten $\pi_i^{\rm tilt}$ for each $T_i$ $(i = 1, 2, ..., 10)$.

Similarly we can compute the transfer matrix and $\pi_j^{\rm trkdir}$ for converting the systematic error in neutrino direction into systematic error in $|\cos\theta_{\rm trk}|$. Once these two sets of numbers, $\pi_i^{\rm trkmm}$ and $\pi_j^{\rm trkdir}$ are ready, we can write the theoretical number of events (including the systematic errors), binned in $|{\rm trkmm}|$ and $\cos(\theta_{\rm trk})$ to be
$$
N^{\rm th}_{ij} = N^0_{ij} \left[ 1 + 0.2*\xi_{\rm norm} + \pi_i^{\rm trkmm} \xi_{\rm trkmm} + \pi_j^{\rm trkdir} \xi_{\rm trkdir}
\right],
$$
where $N^0_{ij}$ is the number of theoretically expected events in the $ij$th bin, according to ${\rm trkmm}$ and $\cos(\theta_{\rm trk})$. This expression for $N^{\rm th}_{ij}$ is used to calculate $\chi^2$.


\begin{thebibliography}{9}


\bibitem{Casper:1990ac} 
D.~Casper {\it et al.},
Phys.\ Rev.\ Lett.\  {\bf 66}, 2561 (1991).
doi:10.1103/PhysRevLett.66.2561

\bibitem{BeckerSzendy:1992hq} 
R.~Becker-Szendy {\it et al.},
Phys.\ Rev.\ D {\bf 46}, 3720 (1992).
doi:10.1103/PhysRevD.46.3720

\bibitem{Hirata:1992ku} 
K.~S.~Hirata {\it et al.} [Kamiokande-II Collaboration],
Phys.\ Lett.\ B {\bf 280}, 146 (1992).
doi:10.1016/0370-2693(92)90788-6

\bibitem{Fukuda:1994mc} 
Y.~Fukuda {\it et al.} [Kamiokande Collaboration],
Phys.\ Lett.\ B {\bf 335}, 237 (1994).
doi:10.1016/0370-2693(94)91420-6

\bibitem{Fukuda:1998mi} 
Y.~Fukuda {\it et al.} [Super-Kamiokande Collaboration],
Phys.\ Rev.\ Lett.\  {\bf 81}, 1562 (1998)
doi:10.1103/PhysRevLett.81.1562
[hep-ex/9807003].

\bibitem{Itow:2001ee} 
Y.~Itow {\it et al.} [T2K Collaboration],
hep-ex/0106019.

\bibitem{Ayres:2007tu} 
D.~S.~Ayres {\it et al.} [NOvA Collaboration],
doi:10.2172/935497

\bibitem{Cleveland:1998nv}
B.~T.~Cleveland, T.~Daily, R.~Davis, Jr., J.~R.~Distel, K.~Lande, C.~K.~Lee, P.~S.~Wildenhain and J.~Ullman,
Astrophys. J. \textbf{496} (1998), 505-526
doi:10.1086/305343

\bibitem{Hampel:1998xg}
W.~Hampel \textit{et al.} [GALLEX],
Phys. Lett. B \textbf{447} (1999), 127-133
doi:10.1016/S0370-2693(98)01579-2

\bibitem{Abdurashitov:2009tn}
J.~N.~Abdurashitov \textit{et al.} [SAGE],
Phys. Rev. C \textbf{80} (2009), 015807
doi:10.1103/PhysRevC.80.015807
[arXiv:0901.2200 [nucl-ex]].

\bibitem{Altmann:2005ix}
M.~Altmann \textit{et al.} [GNO],
Phys. Lett. B \textbf{616} (2005), 174-190
doi:10.1016/j.physletb.2005.04.068
[arXiv:hep-ex/0504037 [hep-ex]].

\bibitem{Abe:2016nxk}
K.~Abe \textit{et al.} [Super-Kamiokande],
Phys. Rev. D \textbf{94} (2016) no.5, 052010
doi:10.1103/PhysRevD.94.052010
[arXiv:1606.07538 [hep-ex]].

\bibitem{Aharmim:2011yq}
B.~Aharmim \textit{et al.} [SNO],
Phys. Rev. C \textbf{87} (2013) no.1, 015502
doi:10.1103/PhysRevC.87.015502
[arXiv:1107.2901 [nucl-ex]].

\bibitem{Steinberger:1990hr}
J.~Steinberger,
Phys. Rept. \textbf{203} (1991), 345-381
doi:10.1016/0370-1573(91)90017-G

\bibitem{Maki:1962mu}
Z.~Maki, M.~Nakagawa and S.~Sakata,
Prog. Theor. Phys. \textbf{28} (1962), 870-880
doi:10.1143/PTP.28.870

\bibitem{Bilenky:1978nj}
S.~M.~Bilenky and B.~Pontecorvo,
Phys. Rept. \textbf{41} (1978), 225-261
doi:10.1016/0370-1573(78)90095-9

\bibitem{Kobayashi:1973fv}
M.~Kobayashi and T.~Maskawa,
Prog. Theor. Phys. \textbf{49} (1973), 652-657
doi:10.1143/PTP.49.652

\bibitem{Kuo:1986sk}
T.~K.~Kuo and J.~T.~Pantaleone,
Phys. Rev. Lett. \textbf{57} (1986), 1805-1808
doi:10.1103/PhysRevLett.57.1805

\bibitem{Apollonio:1997xe}
M.~Apollonio \textit{et al.} [CHOOZ],
Phys. Lett. B \textbf{420} (1998), 397-404
doi:10.1016/S0370-2693(97)01476-7
[arXiv:hep-ex/9711002 [hep-ex]].

\bibitem{Narayan:1997mk}
M.~Narayan, G.~Rajasekaran and S.~U.~Sankar,
Phys. Rev. D \textbf{58} (1998), 031301
doi:10.1103/PhysRevD.58.031301
[arXiv:hep-ph/9712409 [hep-ph]].

\bibitem{Wolfenstein:1977ue} 
L.~Wolfenstein,
Phys.\ Rev.\ D {\bf 17}, 2369 (1978).
doi:10.1103/PhysRevD.17.2369

\bibitem{Wolfenstein:1979ni} 
L.~Wolfenstein,
Phys.\ Rev.\ D {\bf 20}, 2634 (1979).
doi:10.1103/PhysRevD.20.2634

\bibitem{Mikheev:1986wj}
S.~P.~Mikheev and A.~Y.~Smirnov,
Nuovo Cim. C \textbf{9} (1986), 17-26
doi:10.1007/BF02508049

\bibitem{Bahcall:2004fg}
J.~N.~Bahcall and M.~H.~Pinsonneault,
Phys. Rev. Lett. \textbf{92} (2004), 121301
doi:10.1103/PhysRevLett.92.121301
[arXiv:astro-ph/0402114 [astro-ph]].

\bibitem{Fogli:2005cq}
G.~L.~Fogli, E.~Lisi, A.~Marrone and A.~Palazzo,
Prog. Part. Nucl. Phys. \textbf{57} (2006), 742-795
doi:10.1016/j.ppnp.2005.08.002
[arXiv:hep-ph/0506083 [hep-ph]].

\bibitem{Parke:1986jy}
S.~J.~Parke,
Phys. Rev. Lett. \textbf{57} (1986), 1275-1278
doi:10.1103/PhysRevLett.57.1275

\bibitem{Akhmedov:2004ny}
E.~K.~Akhmedov, R.~Johansson, M.~Lindner, T.~Ohlsson and T.~Schwetz,
JHEP \textbf{04} (2004), 078
doi:10.1088/1126-6708/2004/04/078
[arXiv:hep-ph/0402175 [hep-ph]].



\bibitem{Petcov:2005rv} 
S.~T.~Petcov and T.~Schwetz,
Nucl.\ Phys.\ B {\bf 740}, 1 (2006)
doi:10.1016/j.nuclphysb.2006.01.020
[hep-ph/0511277].

\bibitem{Gandhi:2007td} 
R.~Gandhi, P.~Ghoshal, S.~Goswami, P.~Mehta, S.~U.~Sankar and S.~Shalgar,
Phys.\ Rev.\ D {\bf 76}, 073012 (2007)
doi:10.1103/PhysRevD.76.073012
[arXiv:0707.1723 [hep-ph]].

\bibitem{Ghosh:2013mga} 
A.~Ghosh and S.~Choubey,
JHEP {\bf 1310}, 174 (2013)
doi:10.1007/JHEP10(2013)174
[arXiv:1306.1423 [hep-ph]].

\bibitem{Ajmi:2015uda} 
A.~Ajmi, A.~Dev, M.~Nizam, N.~Nayak and S.~Uma Sankar,
J.\ Phys.\ Conf.\ Ser.\  {\bf 888}, no. 1, 012151 (2017)
doi:10.1088/1742-6596/888/1/012151
[arXiv:1510.02350 [physics.ins-det]].

\bibitem{Abe:2017aap} 
K.~Abe {\it et al.} [Super-Kamiokande Collaboration],
Phys.\ Rev.\ D {\bf 97}, no. 7, 072001 (2018)
doi:10.1103/PhysRevD.97.072001
[arXiv:1710.09126 [hep-ex]].


\bibitem{Michael:2006rx} 
D.~G.~Michael {\it et al.} [MINOS Collaboration],
Phys.\ Rev.\ Lett.\  {\bf 97}, 191801 (2006)
doi:10.1103/PhysRevLett.97.191801
[hep-ex/0607088].

\bibitem{Abe:2013fuq} 
K.~Abe {\it et al.} [T2K Collaboration],
Phys.\ Rev.\ Lett.\  {\bf 111}, no. 21, 211803 (2013)
doi:10.1103/PhysRevLett.111.211803
[arXiv:1308.0465 [hep-ex]].

\bibitem{Adamson:2017qqn} 
P.~Adamson {\it et al.} [NOvA Collaboration],
Phys.\ Rev.\ Lett.\  {\bf 118}, no. 15, 151802 (2017)
doi:10.1103/PhysRevLett.118.151802
[arXiv:1701.05891 [hep-ex]].

\bibitem{Cervera:2000kp}
A.~Cervera, A.~Donini, M.~Gavela, J.~Gomez Cadenas, P.~Hernandez, O.~Mena and S.~Rigolin,
Nucl. Phys. B \textbf{579} (2000), 17-55
doi:10.1016/S0550-3213(00)00221-2
[arXiv:hep-ph/0002108 [hep-ph]].

\bibitem{Choubey:2005zy}
S.~Choubey and P.~Roy,
Phys. Rev. D \textbf{73} (2006), 013006
doi:10.1103/PhysRevD.73.013006
[arXiv:hep-ph/0509197 [hep-ph]].

\bibitem{Gandhi:2004bj}
R.~Gandhi, P.~Ghoshal, S.~Goswami, P.~Mehta and S.~U.~Sankar,
Phys. Rev. D \textbf{73} (2006), 053001
doi:10.1103/PhysRevD.73.053001
[arXiv:hep-ph/0411252 [hep-ph]].


\bibitem{Barger:2001yr}
V.~Barger, D.~Marfatia and K.~Whisnant,
Phys. Rev. D \textbf{65} (2002), 073023
doi:10.1103/PhysRevD.65.073023
[arXiv:hep-ph/0112119 [hep-ph]].

\bibitem{Lipari:1999wy} 
P.~Lipari,
Phys.\ Rev.\ D {\bf 61}, 113004 (2000)
doi:10.1103/PhysRevD.61.113004
[hep-ph/9903481].

\bibitem{Narayan:1999ck} 
M.~Narayan and S.~U.~Sankar,
Phys.\ Rev.\ D {\bf 61}, 013003 (2000)
doi:10.1103/PhysRevD.61.013003
[hep-ph/9904302].

\bibitem{Mena:2004sa}
O.~Mena and S.~J.~Parke,
Phys. Rev. D \textbf{70} (2004), 093011
doi:10.1103/PhysRevD.70.093011
[arXiv:hep-ph/0408070 [hep-ph]].

\bibitem{Prakash:2012az}
S.~Prakash, S.~K.~Raut and S.~Sankar,
Phys. Rev. D \textbf{86} (2012), 033012
doi:10.1103/PhysRevD.86.033012
[arXiv:1201.6485 [hep-ph]].

\bibitem{Abe:2018wpn}
K.~Abe \textit{et al.} [T2K],
Phys. Rev. Lett. \textbf{121} (2018) no.17, 171802
doi:10.1103/PhysRevLett.121.171802
[arXiv:1807.07891 [hep-ex]].


\bibitem{Abe:2019vii}
K.~Abe \textit{et al.} [T2K],
Nature \textbf{580} (2020) no.7803, 339-344
doi:10.1038/s41586-020-2177-0
[arXiv:1910.03887 [hep-ex]].

\bibitem{Acero:2019ksn} 
M.~A.~Acero {\it et al.} [NOvA Collaboration],
arXiv:1906.04907 [hep-ex].

\bibitem{HimmelNu2020}
A. Himmel [NOvA],
Presentation at Neutrino 2020, Chicago, June 2020.

\bibitem{Kelly:2020fkv}
K.~J.~Kelly, P.~A.~Machado, S.~J.~Parke, Y.~F.~Perez Gonzalez and R.~Zukanovich-Funchal,
[arXiv:2007.08526 [hep-ph]].

\bibitem{Bharti:2020gnu}
S.~Bharti, U.~Rahaman and S.~Uma Sankar,
[arXiv:2001.08676 [hep-ph]].

\bibitem{Wallraff:2014qka} 
M.~Wallraff and C.~Wiebusch,
Comput.\ Phys.\ Commun.\  {\bf 197}, 185 (2015)
doi:10.1016/j.cpc.2015.07.010
[arXiv:1409.1387 [astro-ph.IM]].

\bibitem{PREM}
A. M. Dziewonski and D. L. Anderson,
Phys. Earth and Planet Int. 25, 297 (1981).




\bibitem{Gandhi:2004md}
R.~Gandhi, P.~Ghoshal, S.~Goswami, P.~Mehta and S.~U.~Sankar,
Phys. Rev. Lett. \textbf{94} (2005), 051801
doi:10.1103/PhysRevLett.94.051801
[arXiv:hep-ph/0408361 [hep-ph]].

\bibitem{Kumar:2017sdq} 
S.~Ahmed {\it et al.} [ICAL Collaboration],
Pramana {\bf 88}, no. 5, 79 (2017)
doi:10.1007/s12043-017-1373-4
[arXiv:1505.07380 [physics.ins-det]].

\bibitem{Esteban:2018azc} 
I.~Esteban, M.~C.~Gonzalez-Garcia, A.~Hernandez-Cabezudo, M.~Maltoni and T.~Schwetz,
JHEP {\bf 1901}, 106 (2019)
doi:10.1007/JHEP01(2019)106
[arXiv:1811.05487 [hep-ph]].


\bibitem{Casper:2002sd} 
D.~Casper,
Nucl.\ Phys.\ Proc.\ Suppl.\  {\bf 112}, 161 (2002)
doi:10.1016/S0920-5632(02)01756-5
[hep-ph/0208030].

\bibitem{Bhattacharya:2014tha} 
K.~Bhattacharya, A.~K.~Pal, G.~Majumder and N.~K.~Mondal,
Comput.\ Phys.\ Commun.\  {\bf 185}, 3259 (2014)
doi:10.1016/j.cpc.2014.09.003
[arXiv:1510.02792 [physics.ins-det]].

\bibitem{GonzalezGarcia:2004wg}
M.~Gonzalez-Garcia and M.~Maltoni,
Phys. Rev. D \textbf{70} (2004), 033010
doi:10.1103/PhysRevD.70.033010
[arXiv:hep-ph/0404085 [hep-ph]].






\end{thebibliography}
\end{document}